\documentclass{PoS}

\usepackage{gensymb}
\usepackage{graphicx}
\usepackage{caption}
\usepackage[rightcaption]{sidecap}

\usepackage{svg}
\usepackage{longtable}
\usepackage{lineno}
\usepackage{verbatim}

\title{A Catalog of Astrophysical Neutrino Candidates for IceCube}

\ShortTitle{Astrophysical Neutrino Catalog}

\author{
The IceCube Collaboration\footnote{For collaboration list, see PoS(ICRC2019) 1177.}\\
{\itshape \href{http://icecube.wisc.edu/collaboration/authors/icrc19_icecube}{http://icecube.wisc.edu/collaboration/authors/icrc19\_icecube}}\\

E-mail: \email{ccardot3@gatech.edu, chujie.chen@icecube.wisc.edu}
}

\abstract{

Multi-messenger astrophysics will enable the discovery of new astrophysical neutrino sources and provide information about the mechanisms that drive these objects. We present a curated online catalog of astrophysical neutrino candidates. 
Whenever single high energy neutrino events, that are publicly available, get published multiple times from various analyses, the catalog records all these changes and highlights the best information. All studies by IceCube that produce astrophysical candidates will be included in our catalog. All information produced by these searches such as time, type, direction, neutrino energy and signalness will be contained in the catalog. The multi-messenger astrophysical community will be able to select neutrinos with certain characteristics, e.g. within a declination range, visualize data for the selected neutrinos, and finally download data in their preferred form to conduct further studies.\\

\vspace{4mm}
{\bfseries Corresponding authors:}
Charles Cardot$^{}$, \speaker{Chujie Chen}\\
{\itshape School of Physics and Center for Relativistic Astrophysics, Georgia Institute of Technology}\\
Atlanta, GA 30332, USA
}

\FullConference{36th International Cosmic Ray Conference -ICRC2019-\\
		July 24th - August 1st, 2019\\
		Madison, WI, U.S.A.}

\begin{document}

\section{Introduction}\label{sec:info}

The IceCube Neutrino Observatory is a $km^3$-scale detector constructed deep within the ice at the geographical South Pole \cite{aartsen2017icecube} to search for high-energy neutrinos from astrophysical sources. Neutrinos are detected by measuring Cherenkov light produced by secondary particles that travel faster than the speed of light in the highly transparent Antarctic ice. Cherenkov light can be used to reconstruct both direction and energy of neutrinos of all flavors. Cherenkov light is measured with over 5000 digital optical modules (DOMs).

Two main types of signals - 'tracks' and 'cascades' - are measured by IceCube \cite{achterberg2006first}. The former are produced by long-range mouns, mostly due to charged current interactions of muon-neutrinos with matter \cite{gupta1989abundance}. Tracks are reconstructed with angular uncertainties of $\sim 0.5^\circ$. Cascades, products of all other neutrino interactions, result from particle showers that are very short, $\sim$20 m, in comparison to the detector dimensions. Typical angular uncertainty for cascades is 10-20$^\circ$ \cite{aartsen2014observation}. High energy tau neutrinos can have more complicated topologies than the ones described. 

The majority of data observed by IceCube are caused by cosmic ray air showers in the upper atmosphere, resulting in a very energetic muon (or muon bundle) that goes through IceCube. Air showers also produce atmospheric neutrinos, which are an irreducible background to astrophysical studies. The challenge is to identify the small portion of astrophysical neutrinos from the atmospheric backgrounds. Background reduction is based on direction, down-going events are dominated by atmospheric muons; energy, astrophysical spectra are harder than background; and quality parameters. 

Although for most events we cannot claim they are definitively from astrophysical origins, astrophysical neutrino candidates can be identified by multiple studies in IceCube. We describe a few below: 

\begin{itemize}
    \item High-energy starting events (HESE): HESE are high-energy neutrino events that interact within the detector volume. This sample is approximately 80\%/20\% cascades/tracks. 
    An anti-coincidence muon veto is used for event selection, identifying neutrino interactions inside IceCube by excluding particles with Cherenkov radiation, notably cosmic-ray muons, entering from outside the detector. At neutrino energies above $\sim$ 100 TeV, the charge threshold of 6000 photoelectrons (PE) significantly reduces the background \cite{aartsen2014observation}. In the original 3-year study, 37 events were detected relative to an background of $8.4 \pm 4.2$ cosmic-ray muon events and $6.6^{+5.9}_{-1.6} $ atmospheric neutrinos \cite{aartsen2014observation}. The latest 7.5-year HESE study focusing a deposited energy range of 60 TeV to 10 PeV has found 103 candidate events \cite{abarbano2019search, wandkowsky_nancy_2018_1301088}.
    
    \item Up-going $\nu_{\mu}$: Up-going muons, due to high energy neutrinos, with estimated muon energies above 200 TeV, are of likely astrophysical origin \cite{aartsen2016observation}. The interaction vertex of muon neutrino events can be outside the detector for the study only covers the Northern Hemisphere where the Earth filters atmospheric muons efficiently. The Earth becomes increasingly impenetrable to neutrinos at the high energies while astrophysical signals follow a harder spectrum than the primary one. The dependence on both the energy and the zenith angle is utilized to form a two-dimensional likelihood fit for selecting signals \cite{PhysRevD.89.062007}. The latest up-going muon sample using eight years of IceCube data is composed of 35 candidate events\cite{haack2017measurement}.
    
    \item Extremely-high-energy (EHE) events: The search for the most energetic events regardless of event topology in IceCube resulting in the first PeV neutrino observation is another source of astrophysical candidates \cite{PhysRevLett.111.021103}. While the high-energy neutrinos targeted by EHE search
    primarily focus on discovery of a cosmogenic neutrino flux\cite{aartsen2016constraints}, it does provide additional astrophysical neutrino events above 1 PeV energies. In a 9-year EHE neutrino search, a Poisson-binned likelihood method is employed and thus, 2 distinct events are identified \cite{syoshida2017Differential}. 
    
    \item The IceCube realtime alert system: Events that pass the online event filtering system are transmitted to the north over a satellite connection. Those events that are likely to be of astrophysical origin decided by several reconstruction algorithms are automatically sent as alerts to Gamma-ray Coordinates Network (GCN\footnote{https://gcn.gsfc.nasa.gov/}) through the Astrophysical Multimessenger Observatory Network (AMON\footnote{http://amon.gravity.psu.edu/}) once a breif summary message has been transmitted to the north \cite{aartsen2017icecubeRealTime, eblaufuss2017realtime}. The realtime system, an important source of astrophysical neutrino candidates, has recently been updated to include additional alert neutrino algorithms \cite{ctung2019realtime}.
\end{itemize}

This proceeding describes the work related to a online catalog that complies all the astrophysical candidate neutrino induced events along with the related experimental resolutions. A given astrophysical candidate may be identified via multiple analyses and/or be presented in multiple IceCube publications. The catalog will highlight the information that is most appropriate for use by the astrophysical community. An example of this is the neutrino alert reported on April 27, 2016. This event was reported in real time via a GCN Notice \cite{GCNnotice160427A}
. This was followed by a more refined reconstruction reported in a GCN Circular \cite{GCNcircular160427A}. Finally, the event is also part of the offline HESE study presented at ICRC 2017 \cite{kopper2017observation} and at ICRC 2019 \cite{hesepos2019}. It would be difficult for a person outside of IceCube to gather the information and choose the suitable one. The catalog already includes data from realtime alerts, and information from other sources will be added in the near future. The catalog can be found \href{https://neutrino-catalog.icecube.aq}{here\footnote{https://neutrino-catalog.icecube.aq}}.

\begin{figure}
\centering
\includegraphics[width=0.9\textwidth]{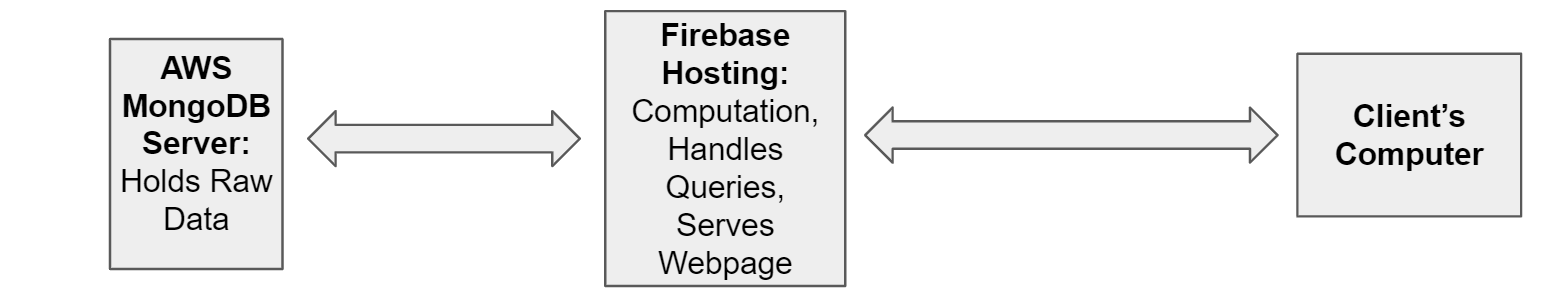}
\caption{A full query possess involves three objects: AWS MongoDB server, Firebase hosting server and a client. Whenever a client makes a request, the Firebase server gives response and communicates bewteen the client and the MongoDB database held on an AWS server.}
\label{fig1}
\end{figure}

\section{Methodology}\label{sec:info}

\subsection{Event Collection}\label{sec:info}

Astrophysical neutrino candidate data have been extracted from GCN Notices and GCN Circulars. In the future, neutrino candidates from other publications will be included. From notices, 10 features including event name, type, time, right ascension, declination and corresponding angular uncertainties are collected into the database (the standard epoch is J2000). When a given event results in multiple alerts, the latest values are stored but only the values preferred by IceCube for public use are displayed at the home page on the web application while the full data set is shown in the event details pages.

\begin{figure}
\centering
\includegraphics[width=1.0\textwidth]{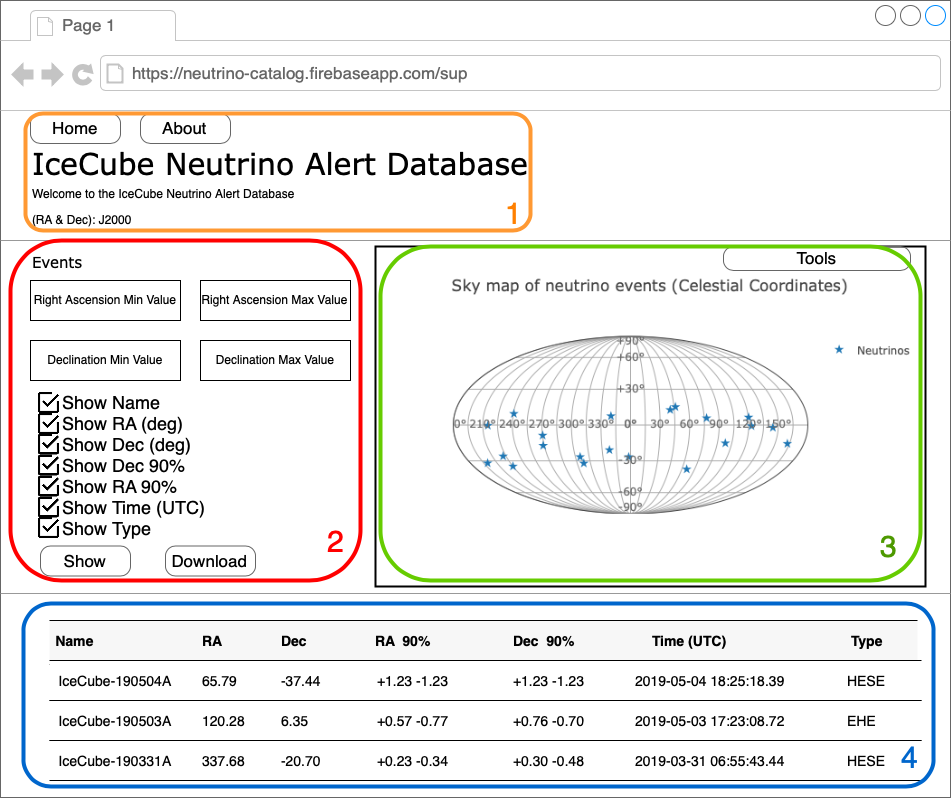}
\caption{A layout of the homepage where the database presents. Area 1: A navigation bar and the title of the page show at the top of the main page. Area 2: Input blanks and check allow users to make queries, update the table below and download events as a CSV file. Area 3: An interactive skymap shows positions of events that are currently shown in the table. Area 4: The table below contains events with values suggested by the IceCube (only a partial of the table is shown).}
\label{fig2}
\end{figure}

\subsection{Database}\label{sec:info}

The database storing astrophysical neutrino candidate events is built upon MongoDB\footnote{https://www.mongodb.com}, an open source database management system (DBMS) developed by MongoDB Inc. MongoDB is a document-oriented DBMS which uses JSON-like schema to manage data entities. The database is structured into subgroups known as collections, in which documents are stored. For the data sets we are dealing with, each document in our collection is defined as a single event that is received by IceCube at a certain time, and different attributes of the event such as event name, type, right ascension and declination are connected with this document. Since the IceCube astrophysical candidate events
are usually reconstructed with different methods and have multiple values, a one-to-many relationship is used to organize this event-to-feature data structure. For the flexibility that MongoDB possesses, arrays are used as the data type of each feature to store various physical values. MongoDB, a NoSQL database which is horizontally scalable, differs from relational DBMS which requires a standard and pre-defined schema, additional features can be added into the database without breaking and redefining relations between different entities. That makes it possible, in the future, for the database to include physical variables that are not already in the present catalog. For example, we currently don't have the event signalness \cite{ialsamarai2017allsky} field in the catalog, but adding this field into the catalog would not cause deficiencies to the original schema. 

\begin{figure}
\centering
\includegraphics[width=0.6\textwidth]{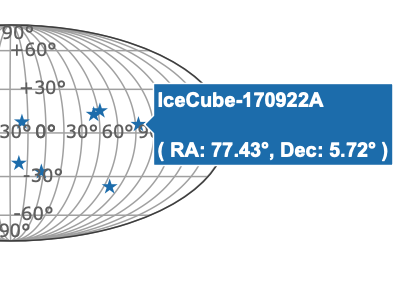}
\caption{The interactive skymap shows all events in the catalog. One can rotate, zoom in, zoom out, select events and save the skymap as an image. A box involving the name and location of an event (IceCube-170922A \cite{icecube2018multimessenger} is used as an example) would appear when one hovers over the mouse on that event.}
\label{fig3}
\end{figure}

The dependence of the current schema on arrays allows us to sort the values of each attribute by reference, according to the position in the array. For example, if two different sources (publications) report an event as having different Right Ascension values, then we can assign index 0 of each attribute array to correspond to the first source and index 1 of each attribute array to correspond to the second source. In this example the Right Ascension attribute array will have different values at the 0 and 1 indexes, organized so that all the same index across each array corresponds to the same source. This allows for all information from each source to be contained within a single document for any given event. We also designate a "default index" for each attribute. For example, if an event has multiple sources, and they report different values for a certain attribute such as Right Ascension, then we can identify which source's Right Ascension value we would prefer to be used by the public. If the default index for Right Ascension for this particular event happens to be 0, then that means that the the value in the zero index of the Right Ascension array is the one we suggest is used by the public. 

Currently the database is hosted on Amazon Web Services (AWS)\footnote{https://aws.amazon.com} for testing purposes and potential cyber security precautions. For now there are only 25 astrophysical neutrino candidates in the database, and the total size of the database is 53.73 KB. Since astrophysical neutrino candidates are rare even though different results are produced for them, the size would not be gigantic and the database will not be a burden for the current server. However, the free AWS can only support no more than 100 concurrent incoming connections, and therefore putting the database on an IceCube server is needed for the future of public use of this database.

\begin{figure}
\centering
\includegraphics[width=0.9\textwidth]{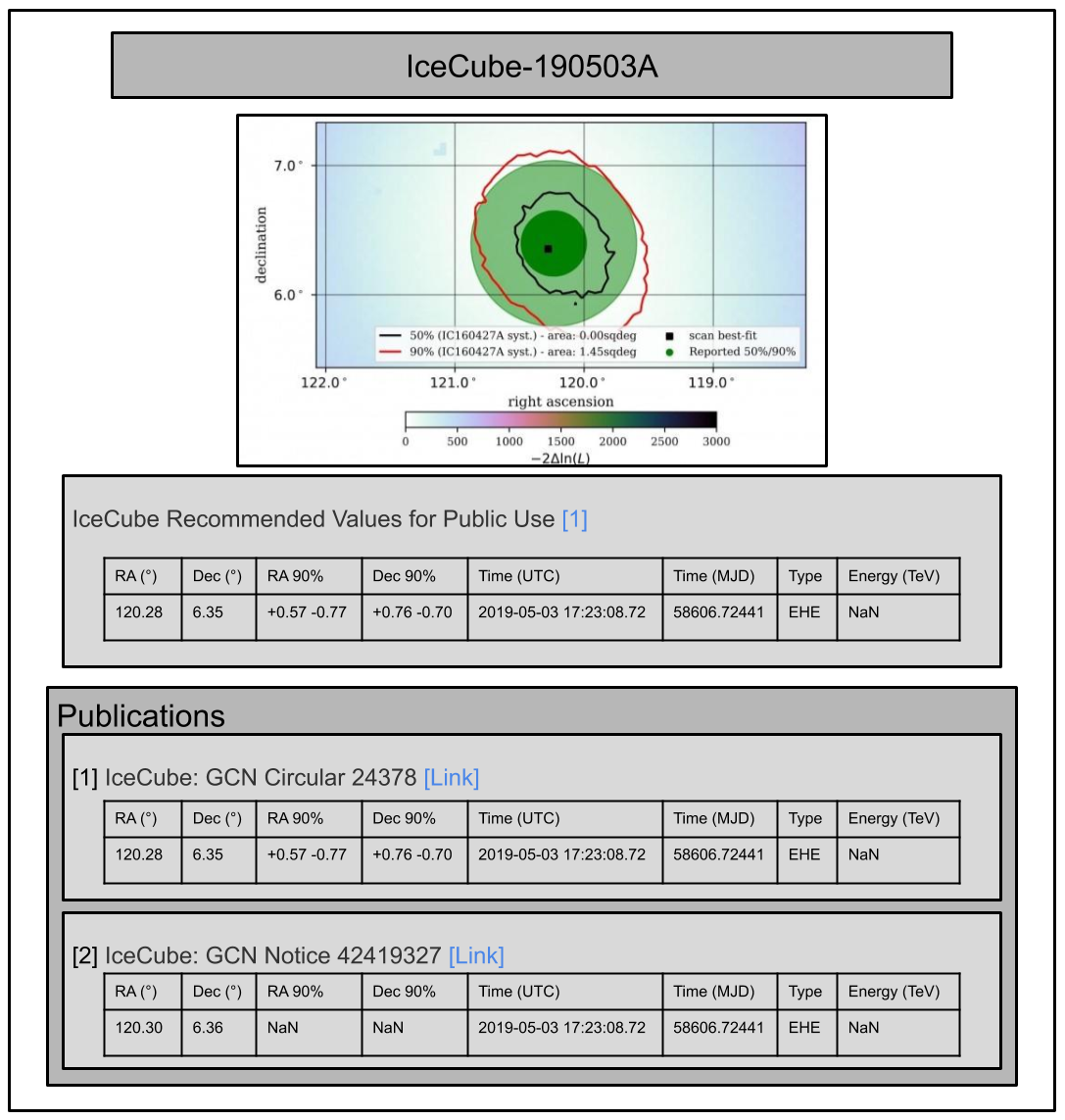}
\caption{An example detail page of IceCube-190503A: An event title is located at the top of the page followed by a likelihood map showing the position of the event. IceCube recommended values are in the first table. All values from published sources are in tables below. For IceCube-190503A, after the GCN notice (the bottom table) was sent out, more sophisticated reconstruction algorithms was applied offline and the GCN circular (the top and the middle tables) was given. The GCN circular is set to be recommended result source.}
\label{fig4}
\end{figure}

\subsection{Web Application}\label{sec:info}

Currently, we are using Firebase\footnote{https://firebase.google.com} to host a website application which acts as a front-end through which users can submit queries to the database. All scripts that handle the computation, sorting, and transmission of query commands, as well as the webpage display, are anchored to a consistently running Firebase server. Each time a client makes a request through the web page, the Firebase server connects to the AWS server which is hosting the MongoDB database and runs the script to pull the data out and send them to the front end - the browser of users (Figure \ref{fig1}).

Node.js\footnote{https://nodejs.org/en/} is utilized as the run-time environment to run JavaScript codes hosted on the host server for its functions can implement scripts in a non-blocking way, so that the client can constantly make queries to produce dynamic web pages. The web application presenting the catalog is written with JavaScript along with visualizing libraries such as D3.js\footnote{Data-Driven Documents, https://d3js.org}, plotly.js\footnote{Plotly JavaScript Open Source Graphing Library, https://plot.ly/javascript/} for the sky map, and EJS\footnote{Embedded JavaScript Templates, https://ejs.co}.

\section{Website}\label{sec:info}

The web application is organized into two types of main pages: the home page and the detail pages. The home page (Figure \ref{fig2}) displays a central table which presents all events in the database along with the IceCube preferred values for each attribute. It also allows for users to select for certain types of events and interact with a visual sky map (Figure \ref{fig3}), which displays all events currently present in the central table. Also on the home page, there is the option to download all data currently presented on the central table into a csv file, so users can use those data to do their own studies. 

Upon clicking on the name of an event, the user is taken to a detail page which gives the user more complete information, organized by publication, and includes comments and images. An example is the sketch in Figure \ref{fig4} showing information about IceCube-190503A. 

\section{Conclusion}\label{sec:refs}

This online catalog already provides the astrophysics community with alert events. In the near future, other astrophysical neutrino events from all IceCube publications will be added into the database. Once the catalog is fully active, it will be continually updated as new events come in. We expect it to be a useful tool for the multi-messenger astrophysical community and more offline studies made by people outside of IceCube, from studying gamma-ray blazar outbursts with HESE event \cite{kadler2016coincidence} to dissecting the region around an IceCube alert \cite{padovani2018dissecting}, will be facilitated by this catalog. 

\bibliographystyle{ICRC}

\bibliography{references}

%

\end{document}